\documentclass[%
 reprint,
 amsmath,amssymb,
 aps,
]{revtex4-2}

\usepackage{graphicx}
\usepackage{dcolumn}
\usepackage{bm}

\usepackage[colorlinks=true, linkcolor=blue, citecolor=blue, urlcolor=blue]{hyperref}

\begin{document}

\title{Parametric resonance and nonlinear dynamics in a coupled double-pendulum system}
\author{Yusheng Niu$^{1,2}$}
\author{Yixian Liu$^{2}$}
\author{Hongyan Fan$^{2,3}$}
\author{Zhenqi Bai$^{2,3}$}
\author{Yichi Zhang$^{2,3}$}%
\email{zhangyichi@sxu.edu.cn}

\affiliation{$^{1}$Sanli Honors College, School of Physics and Electronics Engineering, Shanxi University, Taiyuan, Shanxi 030006, People’s Republic of China\\
	$^{2}$College of Physics and Electronic Engineering, Shanxi University, 030006 Taiyuan, People’s Republic of China\\
	$^{3}$Collaborative Innovation Center of Extreme Optics, Shanxi University, Taiyuan, Shanxi 030006, People’s Republic of China}
\begin{abstract}
Nonlinear dynamics plays a significant role in interdisciplinary fields spanning biology, engineering, mathematics, and physics. Under small-amplitude approximations, certain nonlinear systems can be effectively described by the linear Mathieu equation, which is widely recognized for modeling the response of systems with periodically modulated parameters. Here we investigated a collision-coupled double pendulum system within the framework of Lagrangian mechanics, further explored the nonlinear dynamical characteristics and parametric resonance phenomena at large angular displacements—features that cannot be described by the Mathieu equation alone. Our experiments demonstrate that parametric resonance consistently occurs within a characteristic frequency ratio range ($ \omega /{{\omega }_{0}} $) starting from 2, in agreement with theoretical predictions and numerical simulations. We also find, under periodic driving at moderate frequencies, the system requires initial perturbations to stabilize into periodic states. We propose a novel example in nonlinear dynamics demonstrating large-amplitude parametric resonance phenomena, which also serves as an experimental and theoretical paradigm for exploring classical-quantum correspondences in time crystal research.
\end{abstract}
\maketitle
\textit{Introduction}\textemdash
 The spontaneous breaking of time-translation symmetry \cite{Wilczek2012PRL, Else2016PRL} in Floquet systems is employed to investigate the fractional quantum Hall effect \cite{Kamal2024PRL}. Complicating and perturbations play a crucial role in these systems, serving, for instance, as a tool to evaluate the lifetime of discrete time crystals (DTCs) \cite{Ho2017PRL,Rovny2018PRL}. The integer multiple relationship between the time crystal's period and the driving period \cite{Yao2017PRL} constitutes another key characteristic of discrete time crystals. Experimental platforms include optical lattices, trapped-ion one-dimensional chains and ultracold atoms, among others, all operating at the quantum scale \cite{Wilczek2013PRL,Else2016PRL,Kamal2024PRL,Ho2017PRL,Yao2017PRL,Smits2018PRL,Dobry1990PRB,Zhai2007PRL}. Beyond these quantum systems, Yao's classical discrete time crystal (CDTC) \cite{Yao2020Nature} demonstrates that time-translation symmetry breaking can also emerge in classical regimes through noise-activated first-order phase transitions, sustained by the interplay of interactions and thermal noise in driven-dissipative environments. In classical systems, the Mathieu equation is widely employed in parametrically modulated systems. Remarkably, under specific conditions, the exponentially growing solutions of the Mathieu equation exhibit identical patterns to the subharmonic responses of time crystals \cite{RevModPhys2023}, thereby reinforcing the connection between time crystals and classical dynamics. 

In other classical models, particularly in the study of cosmology \cite{Easther2000PRD,Zibin2001PRD,Paulo2008PRD,Li2025PRD,Eggemeier2024PRD}. the periodic response phenomena induced by such periodic input mechanisms are frequently studied, with a primary focus on parametric resonance. Indeed, parametric resonance also exists in quantum systems \cite{Garc1999PRL,Conforti2016PRL,Lovas2022PRB,Berges2003PRL,Wei2000PRB,Calvo1999PRB}. Beyond cosmic and quantum scales, parametric resonance holds greater practical significance in engineering and mechanical applications, such as enhancing coupling efficiency in voltage-controlled magnetic anisotropy (VCMA) systems \cite{Tomasello2022PRA}, improving signal-to-noise ratios in resonant cavities \cite{Wustmann2013PRB}, and investigating mechanical device microscopic displacement, obstruction and energy input dynamics etc. \cite{Zhao2025PRL, Duka2019,Wang2024,Franco2023,Jerzy2009,Krzysztof2012,Fu2020, Amer2021,Rajasekar2016,Ramachandran2014} System coupling mechanisms exhibit diverse forms, including electromagnetic interactions \cite{Tomasello2022PRA,Chang2017PRB,Wustmann2013PRB,Mu1994PRE,Iqbal2024PRR,Xiao2024PRL,Wei2000PRB,Calvo1999PRB,Tiwari2019PRE,Wang2024} and mechanical stress linkages \cite{Zhao2025PRL,Moshe2019,Ortiz2025PRL,Furst2007PRE,Duka2019,Jerzy2009,Krzysztof2012,Fu2020,Rajasekar2016,Gauld2006}, with most being nonlinear systems. Noting the scarcity of research on systems with intermittent impact coupling and pendulum research often favors the use of the Mathieu equation \cite{Rajasekar2016}. However, this approach is limited to small-angle dynamics and fails to capture the full nonlinear behavior at large angles. Here we selected a collision-coupled double pendulum system for in-depth investigation. Its large-angle displacement component also demonstrates nonlinear characteristics, while the collision-based coupling endows the system with distinctive properties. In contrast to existing studies on collision-coupled double pendulums \cite{Wibowo2023,Wibowo2024,Wibowo20245,Funata2024} and similar models \cite{Gauld2006,Rajasekar2016,Sarkar2023PRE,Choudhury2022PRA}, our work does not focus on detailed kinematic analysis of the system. Instead, we prioritize investigating the global parametric resonance and nonlinear properties. We have developed a large-angle theoretical model and conducted experimental validation, with remarkable consistency between both approaches.

In this work, we derived the collision-coupled double-pendulum model stepwise from Lagrange's equations, followed by numerical simulations and experimental validation. Numerical calculations identified the parametric resonance boundaries of the system and predicted its nonlinear behaviors, laying the groundwork for subsequent experiments. Experimental observations confirmed the existence of a parametric resonance boundary, whose position exhibited amplitude dependence, near a frequency ratio of $ \omega /{{\omega }_{0}} =2 $ (where $ \omega $ is the driving frequency and $ {\omega }_{0} $ the natural frequency). Such phenomena are also observable in the Brownian dynamics of optically trapped water droplets during the transition from overdamped to underdamped oscillations \cite{Leonardo2007PRL}. The parametric resonance occurs at driving frequencies that are integer multiples of the natural frequency, mirroring the characteristic frequency-ratio relationship previously noted in DTCs \cite{Yao2017PRL}. Two studies \cite{Sarkar2023PRE, Rajasekar2016} analogous to our model were also conducted within the Lagrangian framework, employing similar parametric driving. One examined a damped coplanar double pendulum \cite{Sarkar2023PRE}, while the other analyzed a single pendulum \cite{Rajasekar2016}—both representing non-collisional nonlinear systems. Our model, under parametric driving, incorporates both damping and collision effects while also undergoing experimental validation. The works \cite{Sarkar2023PRE,Rajasekar2016} together with our work, collectively affirm the validity of the Lagrangian mechanics-based approach. Our method also overcomes the limitation of the Mathieu equation in similar existing nonlinear studies, which was previously restricted to small-angle approximations, thereby significantly expanding the research scope. Several other nonlinear systems also employ parametric driving methods \cite{Wustmann2013PRB,Mu1994PRE,Salandrino2018PRB,Calvo1999PRB}. In other aspects, nonlinear dynamics has also advanced research on the mammalian cochlea \cite{Alto2020PRR}, nonlinear magneto-optical rotation in rubidium vapor \cite{Put2019PRA} and associating trivalent dendrimer network glasses \cite{Srikanth2013PRE}. 

\begin{figure}[!b]
	\centering
	\includegraphics[width=\columnwidth]{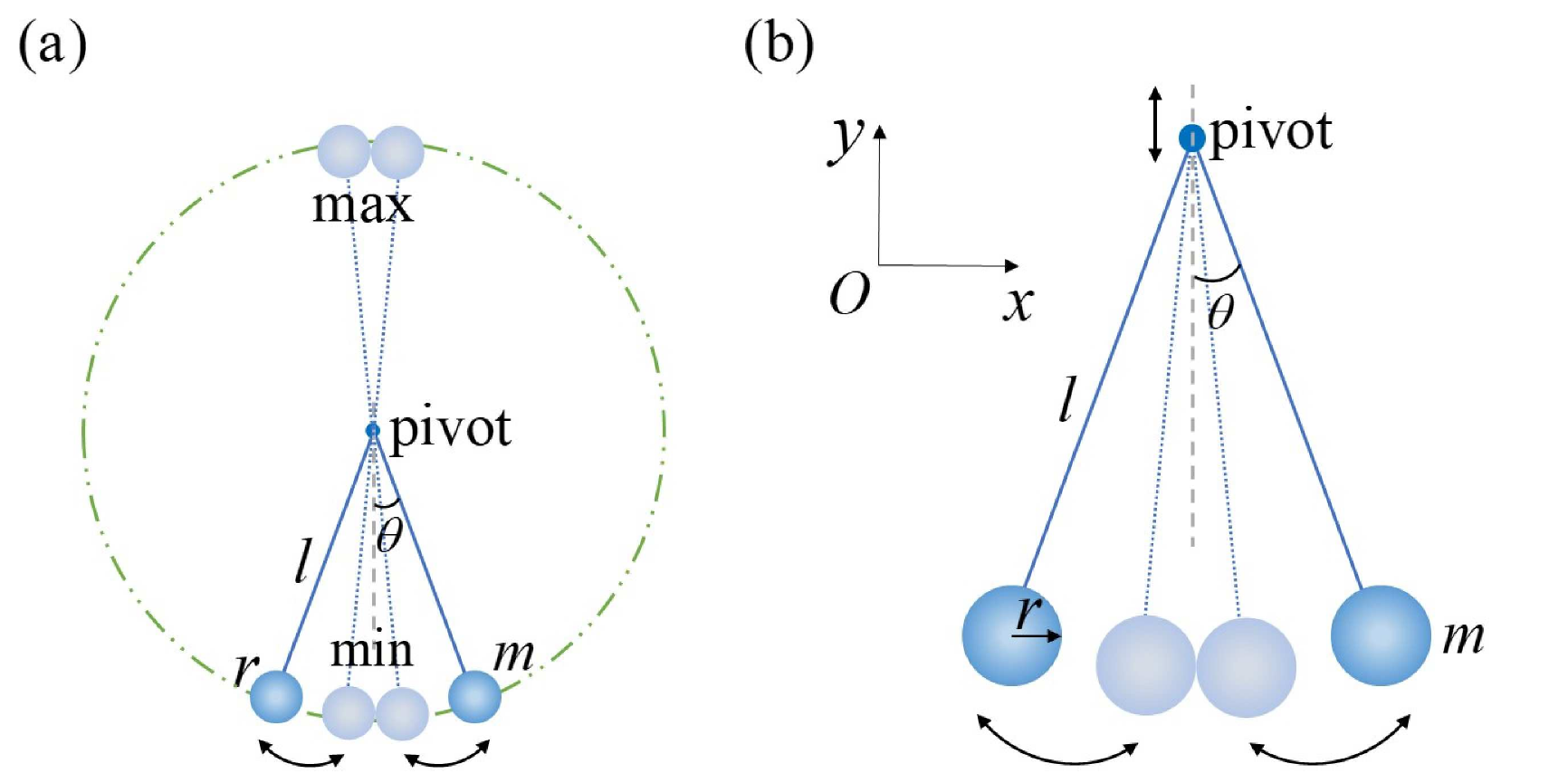}
	\caption{Simplified model diagram of Lato Lato2.0. (a) Ball pivot motion and critical angle. (b) Coordinate system and parameters.}
	\label{fig1}
\end{figure}

The paper is organized as follows. In the Theoretical Framework, we established the model and conducted theoretical derivations, progressively deriving the equations of motion for the two spheres in the system from a single pendulum to a collision-coupled double pendulum using Lagrangian mechanics. In the Numerical calculation, we conducted numerical simulations and identified the boundaries of parametric resonance along with amplitude growth fluctuations in the solutions caused by nonlinearity, subsequently plotting the stability diagram of the system. In the Experimental verification, we conducted two experiments—a single pendulum and a collision-coupled double pendulum — both under parametric driving, and compared the results with numerical simulations. Beyond the strong agreement, we observed that under medium-frequency driving, the system required an initial perturbation to achieve stable parametric resonance and periodic oscillations. This likely compensates for dissipation caused by imperfectly elastic collisions, a phenomenon strikingly similar to the initial perturbations needed to maintain coherence in discrete time crystals.

\textit{Theoretical Framework}\textemdash
We propose three different models.

\begin{figure}[!t]
	\centering
	\includegraphics[width=\columnwidth]{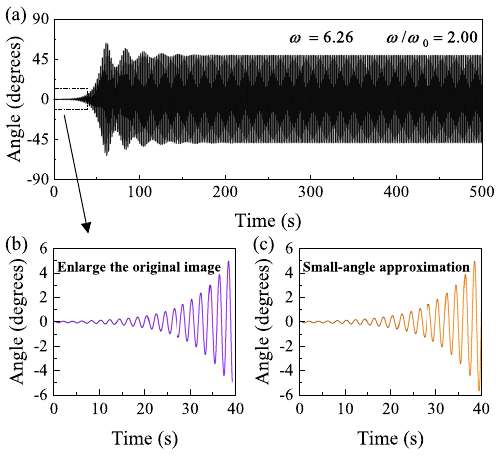}
	\caption{Numerical simulation of the amplitude growth effect under parametric resonance. (a) Full numerical solution of (\ref{eq10}), showing the amplitude growth effect over time. (b) Magnified view of (a) in the time range $0\textendash40\,\mathrm{s}$, highlighting the initial growth phase. (c) Small-angle approximation solution of (\ref{eq10}) for the $0\textendash40\,\mathrm{s}$ range, exhibiting close agreement with  (b). (Parameters: gravitational acceleration $g = 9.8\,\mathrm{m\,s^{-2}}$, pendulum length $l = 1.0\,\mathrm{m}$, ball mass $m = 0.8\,\mathrm{kg}$, damping coefficient $C = 0.05\,\mathrm{kg\,m\,s^{-1}}$, driving amplitude $A_0 = 0.05\,\mathrm{m}$, driving frequency $\omega = 6.26\,\mathrm{rad\,s^{-1}}$, initial angle $\theta_0 = 0.001\,\mathrm{rad}$, initial angular velocity $\dot{\theta}_0 = 0\,\mathrm{rad\,s^{-1}}$.)}
	\label{fig2}
\end{figure}

\textit{(a) Lato Lato 2.0 simplified model.} The simplified model of Lato Lato 2.0 is that the system consists of two light rods of equal length, each with an identical ball suspended from it, and these rods are suspended from a common pivot. The pivot undergoes vertical oscillations under the action of an external driving force, and the balls revolve around the pivot. Under the idealized conditions shown in Fig. \ref{fig2}, the system consists of two identical balls connected by massless rigid rods of equal length, the pendulum plane is strictly perpendicular to the horizontal plane and the oscillations of the pivot are strictly in the vertical direction, thus ensuring that the system is completely symmetrical in the horizontal direction. We establish a 2D Cartesian coordinate system (Fig. \ref{fig1}(b)) where $ l $ denotes the rod length, $ \theta $ is the angle formed by the rod and the negative direction of the $ y $-axis, $ m $ is the mass of the ball, $ r $ is the radius of each ball (non-negligible). The massless rod assumption $ m_{\mathrm{rod}} = 0 $ remains valid even in practical scenarios.

Given the non-negligible ball radius $ r $, when the balls naturally hang vertically and just touch each other, we define the minimum angle $ |\theta_{\min}| = \arcsin\frac{r}{l} $. Similarly, when the balls reach the topmost position of their circular arc, we define the maximum angle $ |\theta_{\max}| = \pi - |\theta_{\min}| $. For convenience, we assume that when the pivot begins to experience external vibrations, the Lato Lato 2.0 device starts oscillating from $ \theta_{\min} $ with an initial angular velocity of zero. Hereafter, $ \theta_{\min} $ and $ \theta_{\max} $ are collectively referred to as the critical angles.

\textit{(b) Static pivot model.} Since the system is completely symmetrical in the horizontal direction, meaning the motion states of the two balls are also completely symmetrical, we can consider the motion of a single ball alone. The motion of a single ball can be regarded as the superposition of the ball's motion relative to the pivot and the motion of the pivot itself. Beginning with the stationary-pivot case (equivalent to a simple pendulum), we derive the undamped equation of motion via Lagrangian mechanics, then extend it to include damping effects like air resistance for the fixed-pivot scenario. Next, we start from the undamped swinging problem (which can also be regarded as a special case of the Lato Lato 2.0 simplified model), using the Lagrangian mechanics method to derive the motion equation of the ball. Further, considering the introduction of damping terms such as air resistance, we obtained the motion equation when the pivot position is fixed and the ball is subject to damping is written.

The right ball has one degree of freedom. In the Lagrangian $ L(q,\dot{q},t) $, we choose the generalized coordinate $ q = \theta $. With the pivot approximately stationary (Fig. \ref{fig1}(b)), the kinetic energy of the right ball is given by
\begin{equation}\label{eq1}
	T = \frac{1}{2}ml^2\dot{\theta}^2,
\end{equation}
where the pivot point serves as both the coordinate origin and zero potential energy reference. The potential energy is
\begin{equation}\label{eq2}
	U = -mgl\cos\theta,
\end{equation}
yielding the Lagrangian:
\[ L = T - U = \frac{1}{2}ml^2\dot{\theta}^2 + mgl\cos\theta.\]
Substituting into the Euler-Lagrange equation
\begin{equation}\label{eq3}
	\frac{d}{dt}\left(\frac{\partial L}{\partial \dot{\theta}}\right) - \frac{\partial L}{\partial \theta} = 0,
\end{equation}
we derive the undamped equation of motion for the right ball
\begin{equation}\label{eq4}
	ml\ddot{\theta} + mg\sin\theta = 0. 
\end{equation}
Ideal conditions would assume the ball's oscillation is unaffected by the surrounding medium, but in reality, such influence is often unavoidable. When an object moves in a medium, its interaction with the medium typically causes a deceleration tendency, and the energy is dissipated in the form of heat. This refers to the heat dissipation within the system. Heat dissipation is often related to the process and involves the conversion of work and energy. Therefore, essentially, this is no longer a purely mechanical problem. Some nonlinear problems are related to thermodynamics \cite{Grasinger2021PRE,Padurariu2012PRB}. In the case of this problem, considering air resistance and introducing a damping term, the motion equation of the ball with damping can still be derived.

For sufficiently small velocities, the frictional force can be expressed as $f = -\alpha \dot{x}$. In this problem, the generalized frictional force (non-conservative force) can be written as
\begin{equation}\label{eq5}
	f = -C\dot{\theta},
\end{equation}
where $C$ is a real-valued constant. 

Unlike the previous section, the system is now non-conservative. Therefore, we must incorporate the non-conservative force term (\ref{eq5}) into the Lagrange equation (\ref{eq3}), yielding
\begin{equation}\label{eq6}
	\frac{d}{dt}\left(\frac{\partial L}{\partial \dot{\theta}}\right) - \frac{\partial L}{\partial \theta} = f.
\end{equation}
(\ref{eq6}) expands to
\[ ml\ddot{\theta} + mgl\sin\theta = -C\dot{\theta},\]
resulting in the damped equation of motion for the right ball
\begin{equation}\label{eq7}
	\ddot{\theta} + \frac{g}{l}\sin\theta + \frac{C}{ml}\dot{\theta} = 0. 
\end{equation}

\textit{(c) Vertical pivot oscillation model.} Observing the motion of the pivot when playing Lato Lato, we assume that the external force applied to the pivot is periodic. Then, we can imagine that under the drive of a sinusoidal force, the pivot will undergo sinusoidal oscillation. Next, by attaching a ground system starting from the Lagrangian quantity, we gradually derive the motion equation of the ball with damping. 

Considering only the right ball with one degree of freedom, we choose the generalized coordinate $ q = \theta $ as shown in Fig. \ref{fig1}(b). The pivot of the suspended ball performs stable sinusoidal oscillation along the y-axis direction. The ball's position in the ground frame at any time can be expressed as the sum of the pivot's position function and the ball's position relative to the pivot. Assuming the pivot's vertical oscillation follows $ y_0 = A_0\cos\omega t $, with $ y_0 = 0 $ as the coordinate origin and zero potential reference point, then the ball's position at time $ t $ is written as
\[ \left\{ 
\begin{array}{l}
	x = l\sin\theta, \\
	y = A_0\cos\omega t - l\cos\theta.
\end{array} 
\right.\]
The kinetic and potential energy terms become
\[ T = \frac{1}{2}m\left(l^2\dot{\theta}^2 + A_0^2\omega^2\sin^2\omega t - 2A_0\omega l\dot{\theta}\sin\omega t\sin\theta\right),\]
\[ U = mg\left(A_0\cos\omega t - l\cos\theta\right).\]
Omitting terms independent of the generalized coordinate and its first derivative, the Lagrangian is written as
\begin{equation}\label{eq8}
	L = \frac{1}{2}ml^2\dot{\theta}^2 - mA_0\omega l\dot{\theta}\sin\omega t\sin\theta + mgl\cos\theta.
\end{equation}
Substituting (\ref{eq8}) into the Lagrange equation (\ref{eq3}) yields the ball's equation of motion
\begin{equation}\label{eq9}
	\ddot{\theta} + \frac{g}{l}\sin\theta - \frac{A_0\omega^2}{l}\cos\omega t\sin\theta = 0.
\end{equation}
To include damping, following (\ref{eq6}), we introduce a damping term to obtain the damped equation of motion
\begin{equation}\label{eq10}
	\ddot{\theta} + \frac{C}{ml}\dot{\theta} + \left(\frac{g}{l} - \frac{A_0\omega^2}{l}\cos\omega t\right)\sin\theta = 0,
\end{equation}
where $\theta$ represents the angle between the light rod and the vertically downward direction at any instant (hereafter referred to as the "angle" or "pendulum angle"), $m$ denotes the mass of a single ball, $C$ is the damping coefficient, $l$ stands for the effective length of the equivalent single rod (hereafter called the "string length"), while ${A}_{0}$ and $\omega$ correspond to the amplitude and frequency of the external vibration respectively (hereafter termed the "driving amplitude" and "driving frequency"). 

\textit{Numerical calculation}\textemdash
Based on the above models, we performed data analysis.

\begin{figure}[!t]
	\centering 
	\includegraphics[width=\columnwidth]{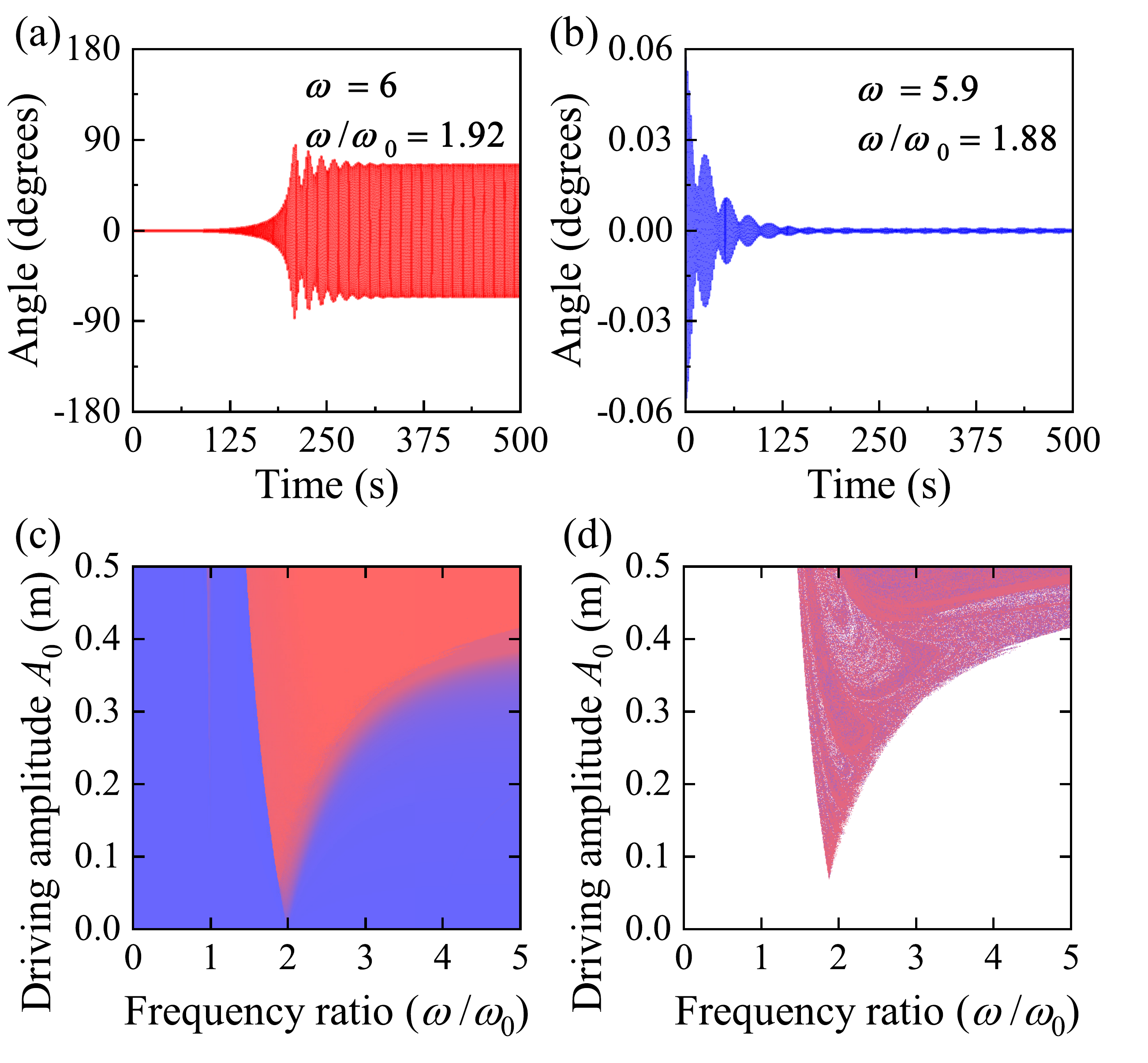}
	\caption{
		Study of parametric resonance in the equation. (a) Parametric resonance occurs with driving frequency $\omega = 6\,\mathrm{rad\,s^{-1}}$. (b) Non-resonant case with $\omega = 5.9\,\mathrm{rad\,s^{-1}}$. Common parameters for (a)-(b): $g = 9.8\,\mathrm{m\,s^{-2}}$, $l = 1.0\,\mathrm{m}$, $m = 0.8\,\mathrm{kg}$, $C = 0.05\,\mathrm{kg\,m\,s^{-1}}$, $A_0 = 0.05\,\mathrm{m}$, $\theta_0 = 0.001\,\mathrm{rad}$, $\dot{\theta}_0 = 0\,\mathrm{rad\,s^{-1}}$.(c) Energy distribution from parameter sweep (red: unstable, blue: stable); (d) Angle-crossing count distribution (dark: unstable, light: stable). Parameters for (c)-(d): $C = 0\,\mathrm{kg\,m\,s^{-1}}$, $r = 0.1\,\mathrm{m}$, $t = 300\,\mathrm{s}$, with stability threshold $|\theta_{\min}|$. Note the order-of-magnitude difference in vertical scales between (a) and (b).
	}
	\label{fig3}
\end{figure}		
\textit{(a) Amplitude growth and parametric resonance.} Numerical solutions of (\ref{eq10}) and (\ref{eq11}) yield the results shown in Fig. \ref{fig2}. To study the solutions in the small-angle regime, the initial angle was not set to the minimum $|\theta_{\min}|$ but to a smaller value $\theta_0 = 0.001\,\mathrm{rad}$. From Fig. \ref{fig2}, it is evident that for $|\theta| \leq 5^\circ$ (corresponding to $0\textendash40\,\mathrm{s}$), the solutions of (\ref{eq10}) and (\ref{eq11}) nearly coincide. As discussed earlier, (\ref{eq11}) closely resembles the Mathieu equation, confirming that the ball's motion in the small-angle regime can indeed be described by the Mathieu equation—a crucial foundation for subsequent analysis. In the large-angle region, the solution of the Mathieu equation exhibits unbounded exponential growth in amplitude, whereas the solution of (\ref{eq10})  displays oscillatory amplitude growth that eventually saturates, reaching a stable amplitude.  Research \cite{Wang2024} indicates that this fluctuation arises because the linear term $\theta$ transitions to the nonlinear $\sin\theta$ as the angle increases, which leads to the stagnation of amplitude changes. The reason for this is the presence of the damping term. This demonstrates that the movement of the ball cannot be fully understood from the Mathieu equation. Moreover, although the damping term is weak, it still has a significant impact on nonlinear systems. It is evident that the introduction of nonlinearity indeed leads to remarkable phenomena. Similar occurrences have been observed in the study of rigidifying curves, where incorporating nonlinear strain in the model results in enhanced bending performance \cite{Al2017PRE}. In the context of the double-slit experiment, certain forms of nonlinearity may generally enable simultaneous measurement of particle trajectories and interference patterns \cite{Valentini1990PRA}. 

However, regardless of the circumstances, as shown in Fig. \ref{fig2}, the solution amplitude of (\ref{eq10}) still exhibits growth when $\omega = 2\omega_0$, suggesting a potential connection to the Mathieu equation. Guided by this observation, we adjusted the driving frequency parameter of (\ref{eq10}) near $\omega = 2\omega_0$ and identified a more significant amplitude growth at $\omega = 1.92\omega_0$ (Fig. \ref{fig3}(a)), which we temporarily call this as "parametric resonance." When the ratio $\omega /{{\omega }_{0}}$ decreases below 1.92, the amplitude undergoes significant attenuation (Fig. \ref{fig3}(b)). This behavior implies the existence of a boundary in the parametric resonance of this nonlinear system, similar to the mobility edge in metal-insulator transitions that separates localized and extended states. Here, such a boundary demarcates stable and unstable motion regimes for the ball in the parameter space. This approach is analogous to studying large-scale nonlinear open quantum mechanics \cite{Childs2016PRA,Roda2024PRR}, where numerical methods are similarly employed to investigate nonlinear systems.

Under this framework, with the system's intrinsic parameters fixed, we conducted a parameter sweep of the external driving vibration, including the driving amplitude ($ A_0 $) and driving frequency ($ \omega $). The sweep range is set as $ A_0 $ from 0 to $ 0.5l $ and $ \omega /{{\omega }_{0}} $ from 0 to 5. The temporal evolution of the system was computed point-by-point throughout $ 300\,\mathrm{s} $. The damping coefficient $ C $ was set to zero to simplify calculations, as its influence can be neglected when the focus is solely on determining whether the amplitude grows, rather than characterizing the specific growth pattern. Figs. \ref{fig3}(c)(d) show the distribution of the system's motion conditions in a two-parameter coordinate system, calculated separately for the average energy at each point within  $ 300\,\mathrm{s}$ and the number of times the angle crosses $ 2\,\mathrm{rad}$. Two criteria determine the stability classification of the ball's motion: higher average energy (labeled as "Criterion 1") or frequent crossings of the $ 2\, \mathrm{rad} $ angle (labeled as "Criterion 2"). The red area in Fig. \ref{fig3}(c) is defined as the non-stable region of the motion, and the blue area as the stable region. Corresponding to Fig. \ref{fig3}(d), the dark areas are the non-stable regions, and the light areas are the stable regions. From Figs. \ref{fig3}(c)(d), it can be seen that regardless of whether Criterion 1 or Criterion 2 is followed, the system naturally presents boundaries between stable and non-stable states, and the left boundary is more distinct than the right boundary. This observation confirms that the phenomenon demonstrated in Figs. \ref{fig3}(a)(b) is universally present in the system. Specifically, when the driving amplitude $ A_0 $ is relatively small and fixed, and when the ratio of driving frequency to natural frequency $ \omega /{{\omega }_{0}} $ is near 2 and crosses a critical value from above, the system abruptly transitions from an unstable state to a stable state. This transition corresponds to the change from amplitude growth to amplitude attenuation observed in Figs. \ref{fig3}(a)(b). Fig. \ref{fig3} conclusively demonstrates the existence of boundaries dividing the parameter plane into distinct regions of stable and unstable motion states for the ball. 

When the ratio $ \omega /{{\omega }_{0}} $ of the driving frequency to the natural frequency is selected as a value within this interval, the system will enter an unstable state. However, if the selected driving amplitude $ A_0 $ is too small, regardless of how $ \omega /{{\omega }_{0}} $ is chosen, the motion of the system will always remain in a stable state. This is the parametric resonance phenomenon of this nonlinear system. By comparing Figs. \ref{fig3}(c)(d), it can be found that beyond the unstable region near $ \omega /{{\omega }_{0}} = 2 $, Fig. \ref{fig3}(c) shows a narrow dark-red band near $ \omega /{{\omega }_{0}} = 1 $, consistent with the Mathieu equation's parametric resonance condition $ \omega \approx 2{{\omega }_{0}} / n $ (for $ n = 1, 2, 3, \dots $). The absence of this feature in Fig. \ref{fig3}(d) indicates that although the motion in this secondary region is more vigorous than its surroundings, it remains significantly less intense than the primary resonance zone near $ \omega /{{\omega }_{0}} = 2 $.

If the region near $\omega /{{\omega }_{0}} = 2$ is referred to as the primary resonance region, then the narrow region can only be considered a secondary resonance region. Naturally, one can imagine that there should also exist sub-secondary resonance regions, sub-sub-secondary resonance regions. However, they are so weak that they are practically undetectable.

\begin{figure}[!b]
	\centering
	\includegraphics[width=\columnwidth]{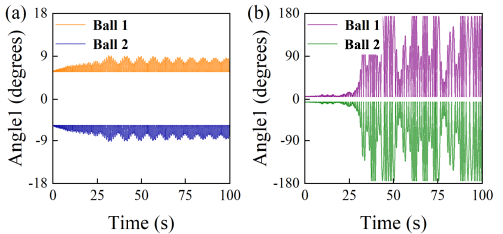}
	\caption{
		Motion of the ball with collision conditions. (a) High-amplitude driving ($A_0 = 0.2\,\mathrm{m}$). (b) Low-amplitude driving ($A_0 = 0.02\,\mathrm{m}$). Common parameters: $g = 9.8\,\mathrm{m\,s^{-2}}$, $l = 1.0\,\mathrm{m}$, $m = 0.8\,\mathrm{kg}$, $r = 0.1\,\mathrm{m}$, $C = 0.02\,\mathrm{kg\,m\,s^{-1}}$, $\omega = 6.26\,\mathrm{rad\,s^{-1}}$, $\theta_0 = |\theta_{\min}|\,\mathrm{rad}$, $\dot{\theta}_0 = 0\,\mathrm{rad\,s^{-1}}$.} 
	\label{fig4}
\end{figure}		
\begin{figure}[!b]
	\centering
	\includegraphics[width=\columnwidth]{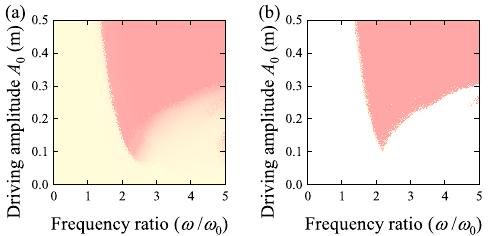}
	\caption{
		Parameter sweep (same specifications as Figs. \ref{fig3}(c)-(d). (a) Energy distribution from parameter sweep (red: unstable, yellow: stable, except near boundaries). (b) Distribution of angle crossing counts from parameter sweep (red: unstable, yellow: stable, except near boundaries).}
	\label{fig5}
\end{figure}

\textit{(b) Collision-coupled double pendulum system.} Both Fig. \ref{fig2} and Fig. \ref{fig3} present numerical solutions of (\ref{eq10}) that deviate from the actual motion of the Lato Lato 2.0 device's dual-ball system in three key aspects. Firstly, the real system restricts angular motion within a semicircular range, preventing the balls from crossing between positive and negative angular regions. Secondly, the amplitude of the balls cannot increase infinitely, as the two balls cannot pass through each other at the highest point of the circular trajectory. Finally, if collisions are taken into account, the evolution of the system over time may change, and Fig. \ref{fig3} may not accurately represent the true parametric resonance behavior of the real Lato Lato 2.0 device.
Based on the above considerations, the conservation of momentum is introduced, that is, when the ball reaches the lowest (or highest) point of the circular trajectory, the angle reaches the critical angle, and the two balls collide. At the moment of the collision, the system's momentum is conserved, and the collision is completely elastic. The collision dynamics \cite{Wibowo2024} are described by
\begin{equation}\label{eq11}
	\left\{ 
	\begin{array}{l}
		v_{1f} = \left( \frac{m_1 - m_2}{m_1 + m_2} \right)v_{1i} + \left( \frac{2m_2}{m_1 + m_2} \right)v_{2i}, \\ 
		v_{2f} = \left( \frac{2m_1}{m_1 + m_2} \right)v_{1i} + \left( \frac{m_2 - m_1}{m_1 + m_2} \right)v_{2i}. 
	\end{array}
	\right.
\end{equation}
The horizontal symmetry conditions $ v_{1i} = -v_{2i} $ and $ m_1 = m_2 $, when substituted into (\ref{eq11}), yield a velocity exchange between the two balls during collisions, causing them to reverse their motion. By adding this physical condition to the numerical calculation of the original equation and imposing constraints on the original solution, we performed the numerical calculation again and obtained the true motion of the two balls in the Lato Lato 2.0 device (as shown in Fig. \ref{fig4}). Following this approach, we developed a GUI program to simulate the system's behavior under fully elastic collision conditions, accurately representing the dynamics of the Lato Lato 2.0 device.
With the inclusion of perfectly elastic collision constraints in (\ref{eq10}), we conducted a parameter sweep following the same specifications as Figs. \ref{fig3}(c)-(d). The sweep results are presented in Fig. \ref{fig5}, demonstrating the modified system behavior under these physical constraints.

By comparing Figs. \ref{fig5}(a)-(b) with Figs. \ref{fig3}(c)-(d), it can be observed that although Fig. \ref{fig5} still retains the same overall structure as Figs. \ref{fig3}(c)-(d), its boundaries have become blurred. Notably, the secondary resonance region no longer appears in Fig. \ref{fig5}. This indicates that the parametric resonance phenomenon in the real Lato Lato 2.0 device is more complex than the phenomenon derived solely from the equations of motion, and a larger driving amplitude is required to observe the resonance effect. In fact, if the driving frequency is properly adjusted—when the system is under these parameter settings (rope length $ l = 1.0 \, \mathrm{m} $, ball mass $ m = 0.8 \, \mathrm{kg} $)—and the driving amplitude $ A_0 $ reaches approximately $ 0.2 \, \mathrm{m} $, the real Lato Lato 2.0 device's balwels can frequently achieve the maximum angular displacement, executing a reciprocating circular trajectory that matches real-world physical behavior. Through numerical simulations, we successfully replicate the actual motion of the Lato Lato 2.0 device's balls. 

\begin{figure}[!t]
	\centering
	\includegraphics[width=\columnwidth]{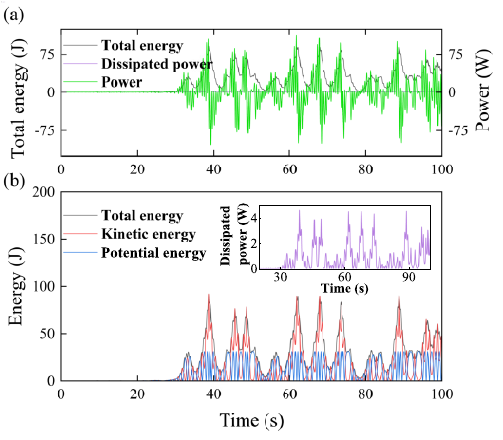}
	\caption{Energy evolution process of the Lato Lato 2.0 system (using the same parameter settings as in Fig. \ref{fig4}). (a)Relationship between input power, dissipated power, and total energy. (b)Relationship between kinetic, potential, and total energy (with inset showing details of dissipated power; note the small magnitude of the vertical scale).}
	\label{fig6}
\end{figure}		

The GUI program described in this article enables the visualization of dynamic behaviors through interactive functions. These include drawing the curves and animation diagrams of the angles of two balls changing with time (the animation supports fast-forward playback), as well as drawing the evolution curve diagram of the system energy over time. 

The numerical program developed in this study calculates the energy evolution of the balls through three primary computational components: (1) kinetic and potential energy based on standard mechanical energy formulas, (2) input power as the product of driving force (inertial force) and velocity in the driving direction, and (3) dissipative power as the product of damping force and velocity in the damping direction. Fig. \ref{fig6} demonstrates the resulting energy evolution process of the Lato Lato 2.0 system under specified parameters.

The total energy evolution of the system exhibits a strong correlation with the input power distribution. When the input power predominantly resides in the positive regime, the total energy demonstrates an increasing trend, whereas a concentration in the negative regime leads to a decreasing trend. Notably, during positive phases, the total energy curve effectively approximates the envelope of the input power curve, a consequence of the input power's substantial magnitude advantage over the irregularly varying dissipative power. Furthermore, the total energy curve also serves as the envelope for both kinetic and potential energy curves, which aligns with fundamental physical principles.

\begin{figure}[!t]
	\centering
	\includegraphics[width=\columnwidth]{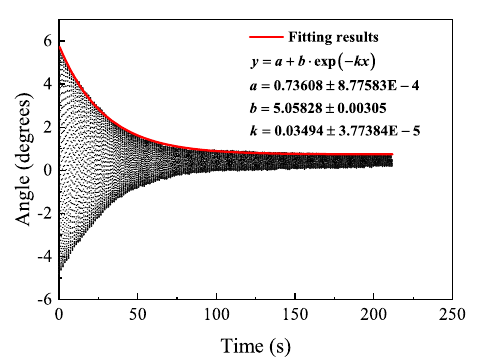}
	\caption{Measurement of the damping coefficient for the single pendulum apparatus. The upper envelope of the $ \theta \mbox{-} t $ curve of the damped vibration is fitted by the power law to obtain the fitting results. The damping coefficient $ C $ is calculated to be $ 0.00274\,\mathrm{kg \cdot m/s} $ from the numerical value of the decay factor $ k $ in the fitting results.}
	\label{fig7}
\end{figure}

\textit{Experimental verification}\textemdash To validate our theoretical analysis and numerical calculations, we designed two experiments: a single pendulum experiment and a coupled double pendulum experiment. Both the simple pendulum experiment and the collision-coupled double pendulum are derived from (\ref{eq10}), with the simple pendulum model serving as the foundational basis for the Lato Lato 2.0 coupled double pendulum system, demonstrating their theoretical continuity. Our experimental setup employed a reciprocating motor with adjustable frequency and amplitude, which converts rotational motion into linear oscillation through a connecting rod mechanism. Under constant current input, we observed the motor's rotary arm performing uniform circular motion, whose projection along the diameter produces near-perfect sinusoidal vibration. Given the rotary arm's significantly shorter length compared to the connecting rod, we confirmed the resulting motion approximates the ideal sinusoidal drive required by our theory.

To minimize out-of-plane oscillations caused by motor rotation, we incorporated a square piston and guide rail system. By attaching the piston vertically below the connecting rod, we effectively constrained the pivot's motion to a single plane while ensuring purely vertical force transmission to our pendulum system.

For the simple pendulum, using thin strings as the connecting material effectively reduces the equivalent rope weight while maintaining clear visibility of the ball's motion trajectory. Even when the string exhibits slight bending as the ball surpasses $90^\circ$, this deformation suffices to determine whether the system is in a high-energy "unstable state" rather than a low-energy "stable state".

\begin{figure}[!t]
	\centering
	\includegraphics[width=\columnwidth]{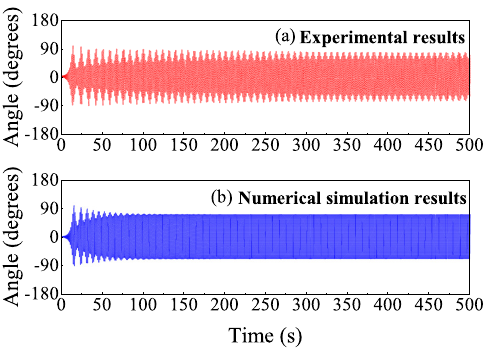}
	\caption{Comparison of numerical and experimental results for pendulum bob dynamics with driving frequency $\omega = 10.09\,\mathrm{rad\,s^{-1}}$ and the frequency ratio $ \omega /{{\omega }_{0}} = 1.942$. (a) Experimental results. (b) Numerical simulation results.}
	\label{fig8}
\end{figure}

\begin{figure}[!b]
	\centering
	\includegraphics[width=\columnwidth]{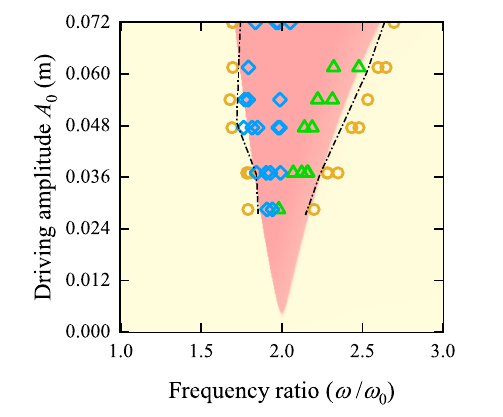}
	\caption{Theoretical and experimental boundaries between stable and unstable regions (red: unstable, yellow: stable, both theoretical; orange dot: no resonance, green triangular dot: weak resonance, blue diamond dot: strong resonance).}
	\label{fig9}
\end{figure}

In the collision-coupled double pendulum apparatus, flexible strings cannot be employed as connecting materials and must be replaced with rigid rods to prevent entanglement during frequent collisions, which rigid rods can effectively mitigate. Furthermore, the use of hollow rods significantly reduces the overall rod weight. The rigid rod design eliminates concerns regarding rod slackening or bending deformation under dynamic loading conditions.

For data acquisition, the experimental videos of pendulum balls and pivot motion were analyzed using a computer, effectively converting raw video footage into quantifiable datasets and visual representations for systematic analysis.

\textit{(a) Simple pendulum experiment.} The experiment employed an equivalent simple pendulum system suspended by three fine strings. Initial measurements of the equivalent length $ l $ yielded the natural frequency $ \omega_0 $, while subsequent mass $ m $ determinations enabled damping coefficient calculations. Experimental results showed excellent agreement with numerical simulations using these parameters, validating the Lagrangian-derived equation of motion (\ref{eq10}).  

Under small-angle approximations ($ \theta \ll 1 $), the pivot-fixed equation (\ref{eq7}) reduces to
\begin{equation}\label{eq12}
\ddot{\theta} + \frac{C}{ml}\dot{\theta} + \frac{g}{l}\theta = 0,
\end{equation} 
with an underdamped solution
\begin{widetext}
\begin{equation}\label{eq13}
\theta = \theta_0 e^{-\frac{C}{2ml}t} \sqrt{1 + \left[ \frac{C}{2ml\sqrt{\frac{g}{l} - \left( \frac{C}{2ml} \right)^2}} \right]^2} \cos\left[ \sqrt{\frac{g}{l} - \left( \frac{C}{2ml} \right)^2}\, t - \arctan \frac{C}{2ml\sqrt{\frac{g}{l} - \left( \frac{C}{2ml} \right)^2}} \right].
\end{equation}
\end{widetext}
The exponential decay factor $-Ct/2ml$ contains the damping coefficient $ C $. Video analysis of free-decay oscillations (Fig. \ref{fig7}) at $ l = 0.3630\,\mathrm{m} $, $ m = 0.10804\,\mathrm{kg} $ yielded $ C = 0.00274\,\mathrm{kg \cdot m/s} $.  

The single pendulum experiment consisted of six sets, with each set having the same driving amplitude $ A_0 $ within the group while differing between groups. Approximately eight trials were conducted per set, with varying driving frequencies measured.  

Figs. \ref{fig8} and \ref{fig9} compare experimental results with numerical calculations under identical parameter settings. Fig. \ref{fig8} shows that the experimental angular growth exhibits exponential increase in the small-angle regime and finite growth in the large-angle regime. As the experiment progresses, the amplitude of angular variation gradually stabilizes. In other words, the system tends toward stability over time, and the experimental results align well with numerical simulations as the system approaches stabilization, validating the theoretical predictions.  

In Fig. \ref{fig9}, the continuous color-mapped background represents numerical results, categorized by energy levels, while the scattered points denote experimental data. Each point corresponds to a single trial, with the pendulum's motion classified into three categories based on the maximum observed angle (since direct energy measurements were impractical): strong resonance (maximum angle reaching $90^\circ$), weak resonance (sustained oscillation with angles above $45^\circ$), and non-resonant regime (small angles at steady state). Notably, cases with angles below $45^\circ$ were rare, indicating sharp parametric resonance transitions. The white dashed boundaries separating stable and unstable experimental states closely match the numerical stability threshold, further confirming the theory's validity.  

\begin{figure}[!b]
	\centering
	\includegraphics[width=\columnwidth]{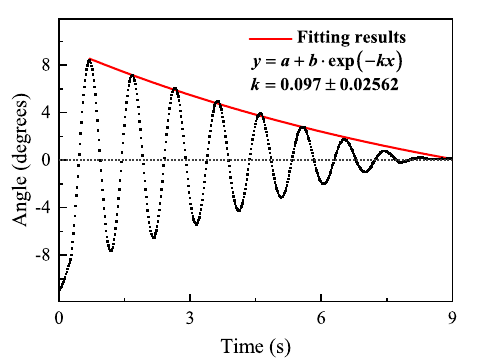}
	\caption{Measurement of the damping coefficient for the collision-coupled double pendulum apparatus. The upper envelope of the $ \theta \mbox{-} t $ curve of the damped vibration is fitted by the power law to obtain the fitting results. The damping coefficient $ C $ is calculated to be $ 0.00506\,\mathrm{kg \cdot m/s} $ from the numerical value of the decay factor $ k $ in the fitting results. }
	\label{fig10}
\end{figure}

\textit{(b) Coupled pendulum experiment.} Using a simplified Lato Lato 2.0 apparatus as the model, we constructed a collision-coupled double pendulum device. The twin lightweight rods were fabricated using bicycle spokes sheathed with hollow thin steel tubes, achieving sufficient rigidity while maintaining the rods' weight at approximately 1/10 of the total system mass. The pendulum tops were connected via low-friction hinges, whose axle was indirectly attached to the base of the square piston. This axle served as the common pivot for both balls, undergoing vertical oscillations driven by a reciprocating motor. Under such vibrational excitation, the balls exhibited mutual collisions, which were systematically observed in this experiment. The damping coefficient was determined through fitting procedures, while the oscillation intensity of the balls was recorded.

\begin{figure}[!t]
	\centering
	\includegraphics[width=\columnwidth]{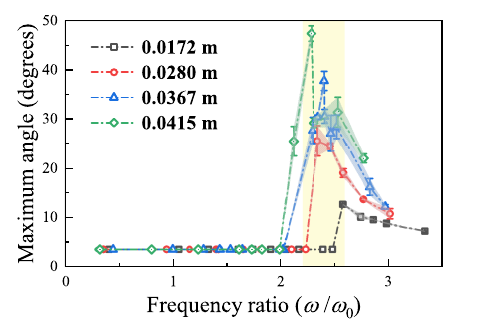}
	\caption{Observation of the parametric resonance phenomenon in the collision-coupled double pendulum system (with the maximum angle of the system after stability as the observable, under different driving frequencies and amplitudes). The peaks in the graph are concentrated in the yellow shaded area and as the driving amplitude increases, the peaks gradually shift to the left.}
	\label{fig11}
\end{figure}
 
Under experimental conditions with string length $ l = 0.2450\,\mathrm{m} $ and ball mass $ m = 0.10644\,\mathrm{kg} $ (as shown in Fig. \ref{fig10}), the measured damping coefficient was approximately $ 0.00506\,\mathrm{kg \cdot m/s} $. As previously discussed, since this study primarily focuses on the oscillation intensity of the coupled double pendulum balls, the influence of damping proves negligible. The obtained damping coefficient is quantitatively reasonable based on order-of-magnitude analysis. Notably, both balls share identical damping sources through the common pivot. The measurement protocol involved suspending one ball while allowing the other to perform underdamped oscillations with a stationary pivot.

The Lato Lato 2.0 coupled double pendulum experiment consisted of four sets, each performing approximately 12 trials with identical driving amplitudes but different driving frequencies. The maximum stabilized angles observed in the videos were plotted as scatter points in Fig. \ref{fig11}, where each point represents one experiment. By connecting points within each set in order of increasing frequency, each curve exhibits a peak occurring at frequency ratios between 2.2 and 2.6. The left side of each peak appears steeper than the right side, and with increasing driving amplitude, these peaks gradually shift leftward. These phenomena precisely correspond to the left and right boundaries observed in numerical calculations (Fig. \ref{fig5}), thereby confirming the theoretical results.

In the experimental observations of the Lato Lato 2.0 apparatus under sinusoidal driving, an initial symmetric perturbation was required to initiate oscillations when the driving frequency remained relatively low. The system consistently evolved toward a stabilized dynamic state (distinct from previously discussed intensity variations), exhibiting remarkable parallels with Floquet time crystal behavior.

Contrary to the theoretical assumption of perfectly elastic collisions at critical angles added to the single pendulum equation of motion (Eq. \ref{eq10}) to complete the coupled double pendulum model, real-world collisions always involve some energy loss. Fortunately, an experimental compensation method was adopted: first adjusting the reciprocating motor to a high-frequency state to excite the balls to a certain angle and velocity (as initial conditions), then readjusting the frequency and studying the subsequent motion of the balls. This preliminary high-frequency excitation provided sufficient initial energy and momentum to compensate for energy and momentum losses due to imperfectly elastic collisions in the materials.

\begin{figure}[!t]
	\centering
	\includegraphics[width=\columnwidth]{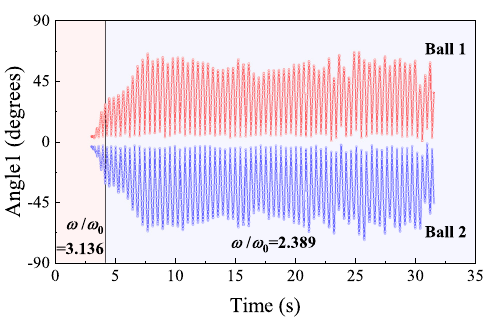}
	\caption{Frequency-stepped energy-momentum injection method (Stage 1: $ \omega /{{\omega }_{0}} = 3.136 $; Stage 2: $ \omega /{{\omega }_{0}} = 2.389 $). Frequency shift at about 4.166s and the transition time less than $ 0.5 \,\mathrm{s} $.}
	\label{fig12}
\end{figure}

Further experiments were conducted within the frequency ratio range of 2.2 to 2.6 in Fig. \ref{fig11}, employing the aforementioned energy-momentum compensation method, yielded the results shown in Fig. \ref{fig12}. The maximum achievable angle of the balls increased significantly, and the experimental values demonstrated excellent agreement with numerical calculations during stable operation, confirming the feasibility of this energy-momentum compensation technique.

\textit{Conclusions}\textemdash
In summary, we systematically investigated the nonlinear dynamics and parametric resonance phenomena in a coupled double pendulum system—Lato Lato 2.0. We first derived the equations of motion using Lagrangian mechanics. Compared to the Mathieu equation, our derived equations remain valid for large angular displacements while converging to the Mathieu equation description \cite{Rajasekar2016} under small-angle approximations, demonstrating the nonlinear nature of our model. Subsequently, we performed numerical simulations of the obtained equations to study the nonlinear effects of amplitude growth and parametric resonance phenomena, introducing perfectly elastic collision conditions to simulate the Lato Lato 2.0 collision-coupled double pendulum system and observe its energy evolution.
 
Based on theoretical and numerical simulations, we designed and conducted experiments, progressively validating from single to coupled pendulums. We observed finite amplitude growth induced by nonlinearity and distinct parametric resonance phenomena related to driving amplitude and frequency, experimentally confirming the theoretical predictions. Additionally, experiments revealed that when the driving frequency was not very high but within the parametric resonance range, the system required an initial external perturbation to initiate oscillations, after which it could reach a periodically varying stable state under periodic external driving—a phenomenon closely resembling Floquet time crystals \cite{Else2016PRL}. The absence of required perturbations at high frequencies might be attributed to strong frequencies causing inherent instrumental disturbances, though the possibility of a new physical phenomenon cannot be excluded. Finally, we proposed a method to compensate for energy and momentum losses during collisions in experiments: a stepwise driving frequency change (first high, then low). This method helps better observe the intensity of ball vibrations in experiments, with the measured maximum stable angles closely matching numerically predicted values. In nonlinear systems, classical and quantum connections often exist \cite{Kolganov2022PRD,ahler1981PRA}. Theories require experimental support \cite{Ma2023PRL}. Even seemingly simple models can encompass profound physics \cite{Maddi2022PRE}.

\textit{Acknowledgments}\textemdash
We thank Yujun Shi and Ximo Wang for discussions. This work is supported by the National Natural Science Foundation of China (No. 61505100), Fundamental Research Program of Shanxi Province (Grant No. 202203021211301), and the Research Project Support by Shanxi Scholarship Council of China (Nos. 2023-028 and 2022-014).

\appendix

\bibliography{apssamp}

\pagebreak
\widetext

\section{Supplemental Materials: Parametric resonance and nonlinear dynamics in a coupled double-pendulum system}

\subsection{Mathieu equation}

For small angular displacements ($|\theta| \leq 5^\circ$), where $\sin\theta \approx \theta$ holds, (\ref{eq10}) can be approximated as
\begin{equation}\label{eq14}
	\ddot{\theta} + \frac{C}{ml}\dot{\theta} + \left(\frac{g}{l} - \frac{A_0\omega^2}{l}\cos\omega t\right)\theta = 0. 
\end{equation}		
The equation undergoes a transition from linear to nonlinear, a phenomenon similarly observed in quantum systems \cite{Bialynicki2001PRA}.

The Mathieu equation holds application value across multiple domains, including laser cooling, free-electron lasers (FEL), ground-state properties and real-time dynamics, quasiperiodically driven nonequilibrium dynamics, and phononic lattices \cite{Maitra2019PRR,Carmesin2020PRR,Bender2020PRR,Periodically2021PRR,Chong2024PRR}. The standard form of the Mathieu equation is expressed as
\[ \frac{d^2x}{dt^2} + \left(a - 2b\cos 2t\right)x = 0,\]
where $ x(t) $ represents the unknown function, $ a $ and $ b $ are constant parameters, and $ \cos 2t $ denotes a periodically varying coefficient with a period of $ \pi $. In various physical systems, a more general parametrized version of the Mathieu equation is commonly employed, which takes the form
\begin{equation}\label{eq15}
	\frac{d^2x}{dt^2} + \left(\omega_0^2 - 2\gamma\omega_0^2\cos(\omega t)\right)x = 0,
\end{equation}
where $ \omega_0 $ is the natural frequency of the system, $ \omega $ represents the angular frequency of external excitation, and $ \gamma $ serves as the excitation strength parameter. Parametric resonance occurs in physical systems when the excitation frequency $ \omega $ satisfies the condition $ \omega \approx 2{{\omega }_{0}}/n $ for $ n = 1, 2, 3, \dots $.

\subsection{Nonlinear phenomena}

\begin{figure}[!b]
	\centering
	\includegraphics[width=0.75\columnwidth]{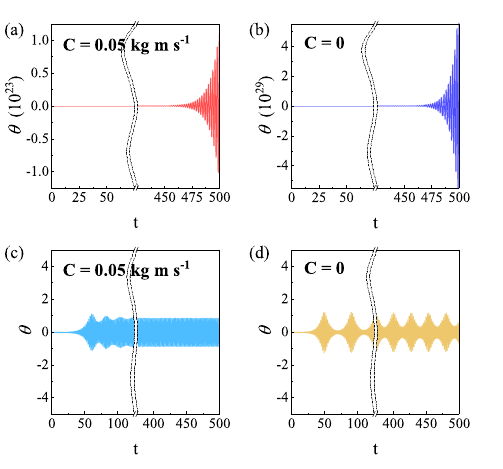}
	\caption{Nonlinear phenomena under parametric resonance (Parameters: gravitational acceleration $g = 9.8\,\mathrm{m\,s^{-2}}$, pendulum length $l = 1.0\,\mathrm{m}$, ball mass $m = 0.8\,\mathrm{kg}$, damping coefficient $C = 0.05\,\mathrm{kg\,m\,s^{-1}}$, driving amplitude $A_0 = 0.05\,\mathrm{m}$, driving frequency $\omega = 6.26\,\mathrm{rad\,s^{-1}}$, initial angle $\theta_0 = 0.001\,\mathrm{rad}$, initial angular velocity $\dot{\theta}_0 = 0\,\mathrm{rad\,s^{-1}}$). The damping coefficients for cases (a) and (c) are $ C = 0.05\,\mathrm{kg\,m\,s^{-1}} $, while for cases (b) and (d) it is 0. Cases (a) and (b) are based on \ref{eq14}, and cases (c) and (d) on (\ref{eq10}).}
	\label{fig13}
\end{figure}

By comparing (\ref{eq14}) and (\ref{eq15}), it is revealed that (\ref{eq11}) has an identical mathematical form to the Mathieu equation (\ref{eq15}). If dissipation is neglected (by setting $ C = 0 $),  then (\ref{eq11}) exactly conforms to the form of (\ref{eq15}). Specifically, in (\ref{eq11}), the system's natural frequency is $ \omega_0 = \sqrt{\frac{g}{l}} $, the external excitation frequency is $ \omega $, and the excitation strength parameter $ \gamma $ can be derived from the factor $ {{A}_{0}}{{\omega }^{2}}/l $. Usually, the damping coefficient C is minute, so the influence of the first-order term in (\ref{eq11}) can be predicted to be very small (Figs. \ref{fig13}(a)(b)). Therefore, for (\ref{eq11}), when the excitation frequency $ \omega $ satisfies $ \omega = 2\omega_0 $, the system will undergo significant parametric resonance, and the amplitude will exhibit exponential growth. The actual result is indeed so (Fig. \ref{fig13}(a)(b)). Naturally, one can speculate that for (\ref{eq10}), if $ \omega_0 = \sqrt{\frac{g}{l}} $ is taken as the natural frequency of the system and $ \omega $ as the angular frequency of external excitation frequency, when $ \omega = 2\omega_0 $, the amplitude will exhibit exponential growth. However, it turns out that this is not the case (Fig. \ref{fig13}(c)(d)). In fact, the amplitude growth becomes bounded and begins to exhibit fluctuations (Fig. \ref{fig13}(c)(d)). This effect arises from a nonlinear term, analogous to how nonlinearity and state-dependence in quantum field theory generate nonlinear quantum mechanics \cite{Kaplan2022PRD,Broz2023PRL,odkiewicz1990PRA} applied to spin freedom \cite{Walsworth1990PRA}. The Fig. \ref{fig13} also reveals that damping exerts a more pronounced influence on motion details in nonlinear systems. But it has no substantial impact on global properties such as the occurrence of parametric resonance. 

\end{document}